The Hα Emission Signature: The Role of Critical Ionization Velocity in Interstellar Space

G.L. Verschuur[1] and J.T. Schmelz[2]

1. verschuur@aol.com
2. USRA, 425 3rd Street SW, Suite 950, Washington, D.C., joan.t.schmelz@gmail.com

ABSTRACT

Gaussian decomposition of HI4PI profiles was done for two regions along high-velocity Complex M, the first with corresponding Hα emission and the second without. In the former, we detected the same Critical Ionization Velocity (CIV) signature found by Verschuur & Schmelz (2022b) for another Hα filament − a line width of 25 km s$^{-1}$. Also discussed are the different CIV signatures found in low-velocity and anomalous-velocity gas. In low-velocity gas, the CIV effect ionizes the atoms and produces the signatures of He (34 km s$^{-1}$), CNO (13.4 km s$^{-1}$), and the metals (around 6 km s$^{-1}$). In anomalous-velocity gas, the CIV effect ionizes the ions and produces the signatures of CNO$^+$ (20 km s$^{-1}$) and/or Hα (25.5 km s$^{-1}$), as well as the CIV signatures of metals around 6 km s$^{-1}$. These distinct CIV signatures indicate that the physical conditions are different in each velocity regime. The particles in the low-velocity gas cloud are initially neutral, the temperatures can be cold (50-100 K), and the abundances can be cosmic. The densities need to be high enough to impose regular recombination. The anomalous-velocity gas, on the other hand, initially contains a significant population of ionized and excited particles. The temperature and abundance values may be similar to those that characterize low-density gas, but the densities must be low enough so that the plasma is essentially collision-less, similar to the conditions in the solar wind.

1. INTRODUCTION

The Critical Ionization Velocity (CIV) concept was first introduced by Alfven (1942, 1954) who proposed that ionization occurs when a plasma and neutral gas are in relative motion in the presence of the magnetic field. Ionization results when the relative velocity of the neutral particle normal to the magnetic field exceeds a critical velocity, $V_c$, defined by

$$V_c = (2eU_i/m_n)^{0.5} \qquad (1)$$

where e is the unit charge, $U_i$ is the ionization potential, and $m_n$ is the mass of the neutral particle. Efficient ionization occurs through the intermediate role of the two-stream plasma instability (Peratt 2015) that transfers energy to seed electrons, which, in the case of interstellar matter, exist everywhere. Their resulting distribution function is pseudo- Maxwellian at low energies with a high energy tail containing electrons with energies that exceed the ionization energy of the neutrals. Candidates for this process have been proposed by many authors (Danielsson 1973, Sherman 1973, Raadu 1978; Galeev 1981; Abe 1984; Abe & Machida 1985; Papadopoulos 1985). The CIV effect has been observed in a series of laboratory experiments (Fahleson 1961; Axnäs 1978), in plasma gun devices (Danielsson 1970), in the terrestrial magnetosphere (Haerendel 1982), in cometary environments (Galeev et al. 1986; Formisano et al. 1982), and in the interaction between the solar wind and interstellar gas (Petelski et al. 1980). Please see Lai (2001) for a comprehensive review.

Observational evidence of the CIV signature in interstellar space was first reported by Verschuur & Peratt (1999) and Peratt & Verschuur (2000). Verschuur & Schmelz (2010) recognized that interstellar neutral hydrogen (HI) profiles exhibit a widespread broad underlying component with a velocity width of 34 km s$^{-1}$. This width would appear to correspond to a temperature high enough to ionize the gas so it could not contribute to the λ21-cm profile.

However, given that the CIV for helium is 34.3 km s$^{-1}$, this raised the intriguing possibility that HI line width of this beyond-thermal component may result from the CIV effect. Peratt (2015) shows that electrons spiraling about B imparts a thermal plasma distribution to the neutral gas with velocity components in all three spatial directions. This phenomenon has been investigated by Machida & Goertz (1988). Given that the motion of neutral hydrogen atoms is coupled to the electrons, it is easy to show that a random distribution of velocities between the limits of ± 34 km s$^{-1}$ will produce a Gaussian line width of this order.

The CIV interpretation of HI linewidths was not immediately accepted by the community, who adopted a more traditional explanation where the line width was set by the gas temperature modulated by turbulence. This view appeared to be supported by a distribution of Gaussian line widths for low-velocity gas from Nidever et al. (2008). Their distribution was not sharply peaked at 34 km s$^{-1}$ as is expected if the CIV effect were operating in interstellar space but, rather, had a long tail of very broad line components extending out to hundreds of km s$^{-1}$. Since Nidever et al. (2008) and Verschuur & Schmelz (2010) both used data from the Leiden/Argentine/Bonn (LAB; Kalberla et al. 2005) all-sky survey, it was not clear why the results were different. Verschuur & Schmelz (2022a) subsequently showed that the discrepancies arise when a Gauss-fitting algorithm finds component fits that are clearly incorrect. They showed that when the problems are recognized, the reality of a family of line widths of order 34 km s$^{-1}$ is clearly revealed.

Verschuur & Schmelz (2022b) found evidence for an Hα CIV signature. They studied the detailed structure of an Hα filament observed by Ogden & Reynolds (1985) using λ21-cm emission from the Effelsberg-Bonn HI Survey (EBHIS; Winkel et al. 2016). The relatively simple morphology and isolated location in both velocity and position make this filament an excellent candidate for study since there was no confusion from emission at unrelated velocities or background features. The filament appears to be part of a complex gas structures flowing out of the Perseus arm and is located of order 1 kpc from the Galactic plane. Neither traditional photoionization nor shocks seems to be responsible for the energy required to explain the presence of Hα at this location. Gaussian analysis of both the Hα and HI profiles revealed a line width of about 25 km s$^{-1}$. After emission of the Hα photon the hydrogen atoms find themselves in the n=2 energy level which has an ionization potential of 3.2 eV. According to Eqtn. 1 the CIV is then 25.5 km s$^{-1}$. This close agreement with the observed line width data appears significant. The presence of Hα emission in any given area of sky leads to an expectation of the CIV signature in the HI emission profiles, and the corollary is that if analysis of HI profiles in an area reveals a preponderance of 25 km s$^{-1}$ wide components, then Hα emission is likely to be present. These results provide additional evidence that the CIV effect is operating in interstellar space, and this may account for the previously unknown excitation mechanism responsible for the Hα emission (Ogden & Reynolds 1985; Haffner et al. 1999).

Because it is difficult to observe helium directly in astrophysical environments, the presence of the 34 km s$^{-1}$ wide neutral hydrogen emission line component opens a new window into understanding the composition of interstellar matter. Hα surveys, on the other hand, suggest a new experiment to explore the possible role of the CIV effect in interstellar space. In this paper, we expand the Hα-HI joint analysis to regions along high-velocity Complex M. In §2, we compare neutral hydrogen features with and without corresponding Hα emission. In the former, we expect to detect the CIV signature found by Verschuur & Schmelz (2022b) − a line width of 25 km s$^{-1}$. In §3, we discuss a possible explanation for our findings, and in §4, we summarize our conclusions.

2. ANALYSIS

The λ21-cm Galactic neutral atomic hydrogen data from the HI4PI Collaboration (Ben Bekhti et al. 2016) combines the northern hemisphere data from EBHIS and the southern hemisphere data from the third revision of the Galactic All-Sky Survey (GASS ; McClure-Griffiths et al. 2009). HI4PI has an angular resolution of $\Theta_{FWHM} = 16.'2$, a sensitivity of σ = 43 mK, and full spatial sampling, 5′ in both galactic longitude (l) and latitude (b). The Hα observations were obtained on a grid of 1° in size with an effective beam width of 0.°8 (Ogden & Reynolds 1985).

Fig. 1 shows a map of l versus b for the region of Complex M that motivated this study. The HI4PI data contours at −80 km s$^{-1}$ are superposed on the Hα image in color for the same velocity. Some HI contours clearly coincide with patches of Hα while others do not. Like the filament studied by Verschuur & Schmelz (2022b), the isolated location in both velocity and position makes this segment of Complex M an excellent candidate for study without being confused by emission at unrelated velocities or background structure. It is also challenging to explain how traditional photoionization might be responsible for the energy required to account for the presence of Hα in this area of sky, given its high galactic latitude and anomalous velocities.

The two areas selected for detailed Gaussian analysis are shown in Fig. 2. These are designated the "Hα region" and the "non-Hα region" throughout the paper. The vertical dashed lines in each panel indicate the positions of the HI profile decomposition. Here, we make use of the semi-automated Gaussian fitting method described by Verschuur (2004) and Verschuur & Schmelz (2022a), which employs the Microsoft Excel SOLVER algorithm. (SOLVER uses the generalized reduced gradient (GRG2) nonlinear optimization code developed by L. Lasdon, University of Texas at Austin, and A. Waren, Cleveland State University. References to papers published on this method may be found at http:// www.optimalmethods.com.)  When done carefully and correctly, the results of Gaussian analysis can provide insights into the physical conditions present in the interstellar medium, but please see Verschuur & Schmelz (2022a) for a discussion of the limitations of Gaussian decomposition, which also apply here.

We analyzed 50 HI4PI profiles in the Hα region and 61 profiles in the non-Hα region resulting in 558 & 582 Gaussian components covering the full velocity range of the observed HI emission.

The number of components with center velocities between -70 & -90 km s$^{-1}$ in the two areas are 100 and 104, respectively. (Note: Only components with brightness temperature ≥ 0.1K and column densities ≥ $1.0 \times 10^{18}$ cm$^{-2}$ were included in the analysis to avoid confusion generated by noise.) Fig. 3 shows a typical spectrum for (a) the Hα region and (b) the non-Hα region. Although the overall profiles appear complex and as many as eight individual components are required to achieve a good fit across the full velocity range, it is only the relatively isolated, left-most component that is the target of the present study. These are indicated by a thick red line in the figure and clearly represent extremely good fits to the highest negative-velocity components of the profiles. The numbers beside the dominant features are their line widths in km s$^{-1}$. The components plotted with dark lines are members of the 34 km s$^{-1}$ family of components identified by Verschuur & Schmelz (2022b).

Fig. 4 a & b are the line width histograms for components with center velocities between -70 and -90 km s$^{-1}$ for the Hα region and non-Hα region, respectively. Each distribution shows the expected narrow components, but the new and most interesting feature is the dominant, wider peak at 25.9 ± 1.6 km s$^{-1}$ for panel (a) and 20.7 ± 1.5 km s$^{-1}$ for panel (b). The results in Fig. 4a confirm our prediction − the peak is consistent with the CIV for excited hydrogen in the n=2 energy level of 25.5 km s$^{-1}$, in agreement with the Hα CIV signature found by Verschuur & Schmelz (2022b). The dominant peak in Fig. 4b is distinct and statistically distinguishable from the Hα CIV signature. In fact, "forcing" the SOLVER algorithm to use a component this broad results in a visibly bad fit to the data. This lower line width is consistent with the cluster of CIVs for the ionized components of several of the most abundant elements in the interstellar medium, $C^+$, $N^+$, and $O^+$, which is 20.0 ± 0.4 km s$^{-1}$. These four elements are the primary members of Band II in the list of CIV values shown in Table 1 of Peratt & Verschuur (2000). For convenience these will be referred to as the CNO components. We discuss the physical implications of these results in the next section, but first we need to clarify the width of the histogram peaks themselves, which are related to receiver noise.

Verschuur & Schmelz (2010) investigated the effects of noise on the Gaussian decomposition analysis. They generated 150 synthetic profiles, each with a peak temperature of 1.6 K, center velocity 0 km s$^{-1}$, and line width 34 km s$^{-1}$, plus one or two additional components with different strengths and velocities. A random number generator was then used to add varying amounts of noise with amplitudes between 0.05 and 0.1 times the peak to each profile. The results were then run through the same SOLVER algorithm used here. The histogram of line widths (their Fig. 10) clearly shows the three input components, but there is a spread in the results. In other words, the exact value of the input line width is not recovered for every noisy spectrum, but overall, SOLVER is able to recover the line width for the ensemble. These results are similar to those seen here in Fig. 4 and help us understand the histogram spread around the peak values of 25 and 20 km s$^{-1}$ in panels (a) and (b), respectively. This spread is not only recognized as, but expected for, any Gaussian decomposition of spectra involving noise. It is the histogram peak that we want to relate to the physical parameters that may be operating in interstellar space. The spread, on the other hand, is an indication of the noise level.

3. DISCUSSION

To understand the implications of the previous section it is important to step back and put these results into context with the outcomes of previous CIV analysis for the interstellar medium. The original insight was that Gaussian analysis revealed several families of components with line widths that corresponded to the CIV s of helium (34.3 km s$^{-1}$), CNO (13.4 ± 0.7 km s$^{-1}$), and metals from 5 to 8 km s$^{-1}$ (Verschuur & Peratt 1999; Peratt & Verschuur 2000; Verschuur & Schmelz 2010). The second insight, described in more detail below, is that low-velocity gas, i.e., nearby gas in the Galactic disk, shows different CIV signatures than anomalous velocity gas, i.e., gas that does not conform to the differential rotation of the Galaxy.

Fig. 5 helps us visualize this context. Panel (a) plots the Gaussian line width distribution for an elongated filament in the Southern Galactic sky studied by Verschuur et al. (2018) separated according to the second insight described above. The low-velocity gas, from -25 to +25 km s$^{-1}$, was decomposed into 584 Gaussians components and is shown as dark bars; the anomalous-velocity gas, with velocities less than -25 km s$^{-1}$, resulted in the 167 components shown as red bars. These data are based on profile samples along six equally spaced cuts across the width and over a 3° length of the filament. The dark bars in Fig. 5a show peaks at approximately 34, 14, and 6 km s$^{-1}$, a pattern found for other low-velocity profiles described by Verschuur & Schmelz (2022a) and references therein. In contrast, the red bars for the anomalous velocity gas show components from 4 to 10 km s$^{-1}$, which we attribute to metals, as well as components between 18 and 28 km s$^{-1}$. The WHAM survey data reveals patchy filamentary Hα emission over this area.

Fig. 5b shows similar results, but for a different direction − toward a segment of the high-velocity cloud MI − where a uniform grid of profiles separated by 0.°2 in both longitude and latitude was analyzed. The dark bars depict 597 low-velocity components, and the red bars depict the 654 components with velocities between -90 & -120 km s$^{-1}$. The distinction between peaks at 14 and 6 km s$^{-1}$ seen in Fig. 5a is not so clear here, but the red bars around 25 km s$^{-1}$ may be the CIV Hα signature. This would be consistent with the Hα emission associated with MI that has been reported by Tufte et al. (1998).

Putting together all the points raised above, the list below summarizes these findings and their relation to the CIV effect. The next step is to interpret these results and understand what they appear to tell is about the conditions in the interstellar medium. Thus, we suggest the following:

Low-Velocity Gas

- Metals ionize via the CIV effect and produce CIV signatures of a few km s$^{-1}$; most ions recombine and are available for further ionization; some may remain, contributing electrons and ions to the interstellar medium.
- CNO ionizes via the CIV effect and produces the 14 km s$^{-1}$ CIV signature; most ions recombine and are available for further ionization; some may remain, contributing electrons and ions to the interstellar medium.
- He ionizes via the CIV effect and produces the 34 km s$^{-1}$ CIV signature; most ions recombine and are available for further ionization; some may remain, contributing

- electrons and ions to the interstellar medium. Ionization of He$^+$ would, in turn, produce alpha-particles and electrons, seeds for cosmic rays.
- H ionizes via the CIV effect but produces no CIV signature in 21-cm profiles; most protons recombine and are available for further ionization; some may remain, contributing electrons and protons to the interstellar medium.

Anomalous-Velocity Gas

- Metals are ionized via any mechanism but might not recombine because of super-low densities; ions are available for a second ionization via the CIV effect and produce CIV signatures of a few km/s; the low CIV signatures for excited states are indistinguishable from each other.
- CNO is ionized via any mechanism but might not recombine because of super-low densities; CNO ions are available for a second ionization via the CIV effect and produce the 20 km s$^{-1}$ CIV signature; the low CIV signatures for excited states are indistinguishable from each other.
- He is ionized via any mechanism but might not recombine because of super-low densities; Helium ions are available for a second ionization via the CIV effect and produce a 51 km s$^{-1}$ CIV signature, which is unobservable; excited state transitions are not common, so are not likely to be observable.
- H is ionized via any mechanism and the excited state Hα produces a CIV signature at 25 km s$^{-1}$.

A key to understanding these phenomena is the fundamentally different CIV signatures in the low- and anomalous- velocity gas seen in Fig. 5. Let us consider the low-velocity gas first. A "cloud" of neutral hydrogen with cosmic abundances interacts with a stream of plasma. Let us also make the simplifying assumption that, although this cloud is in the Galactic disk, it is far from regions where stars are either forming or dying. If the velocities normal to the magnetic field are of the right magnitude, the interaction between the cloud and the stream triggers the CIV effect, ionizes the atoms, and produces the signatures of He (34 km s$^{-1}$), CNO (14 km s$^{-1}$), and the metals (around 6 km s$^{-1}$), the dark bar peaks in Fig. 5. (Note: the CIV for hydrogen itself is 51 km s$^{-1}$, but after ionization there is no HI left to contribute to the λ21-cm profiles.) Also important is what we do not see − the 21 or 25 km s$^{-1}$ components indicating ionized or excited particles undergoing the CIV effect. This tells us that recombination acts quickly, returning at least most of the ions to their neutral form where they may be ionized again if the conditions are right. This continuous process is different than the impulsive events that are studied in laboratory experiments; otherwise, the CIV process would remove all the neutral gas until the plasma stream faded away.

What do these results tell us about the interstellar medium? The first is that the hydrogen gas in the cloud can be cold, 50-100 K, and still produce a λ21-cm profile with broad Gaussian components that are traditionally attributed to higher temperatures. The second is that the densities are sufficiently high so collisions occur quickly enough to neutralize the ions before that can undergo further ionization. Although we are not measuring abundances directly here, it is noteworthy that the CIV signatures are detected for the most abundant elements. Pervasive turbulence keeps the interstellar matter moving and interacting, even if the clouds are cold. Finally, the magnetic field is vital, perhaps even as important as it is in influencing and controlling the processes that play a role in interplanetary space.

Let us next consider an anomalous-velocity cloud. The CIV signatures seen in Fig. 5 tell us that at least some of these clouds contain a substantial population of ions that could have been formed in any number of ways − supernovae, shocks, photoionization, CIV effect, etc. When this cloud moves with velocities normal to the magnetic field that are of the right magnitude, the interaction triggers the CIV effect, ionizes the ions, and produces the signatures of $C^+$, $N^+$ & $O^+$ (20 km s$^{-1}$) and/or Hα (25.5 km s$^{-1}$), the red bar peaks in Fig. 5. (Note: the CIV for ionized helium is 51 km s$^{-1}$, which would not be detected in the λ21-cm profiles because neutral hydrogen is also ionized at this velocity.)

Gaussian line widths of 20 km s$^{-1}$ would traditionally be attributed to a warm neutral or ionized medium with a temperature of about 8,000 K (Payne at al. 1980), which might be in pressure balance with both colder and hotter interstellar components. The old "textbook" explanation for the source of this energy would be HII regions, where emission nebulae are created when young, massive stars ionize nearby gas with high-energy UV radiation. Recent results paint a more complex picture. Evidence suggests that heating of low-density regions of the interstellar medium is dominated by sources other than photoionization (Haffner et al. 1999; Reynolds et al. 1999).

Even though there is no obvious source of photoionization for either the Hα region or the non-Hα region, it is important to consider temperature as the explanation of the Gaussian line widths seen in Fig. 4. Although it is true that the peak width of 25 km s$^{-1}$ in Fig. 4a could result from 13,700 K gas with enough energy to excite or ionize the hydrogen and produce the observed Hα photons. The exact same temperature argument could be made regarding the peak in Fig. 4b, the region where no Hα is observed. So, if the line widths really reflected the gas temperature, both regions should be emitting Hα photons, and this is clearly not the case.

These results reveal additional properties of the interstellar medium. Because a substantial population of ions must be present, the densities in these clouds may be low enough that recombination is inefficient. The solar wind provides a good analogy. Here, atoms are ionized in the million-degree corona but then flow quickly into the low-density environment of interplanetary space. The nearly collision-less plasma allows the particles to remain ionized as they travel through the solar system. This analogy may not only explain the CIV signatures in Fig. 5, but also account for the presence of ions in interstellar space far from known sources of ionizing radiation.

Finally, as Figs. 2 & 3 show, we observe either 25 or 21 km s$^{-1}$ lines in a certain direction, but not both. The choice could simply be related to the component of the velocity of the plasma cloud normal to the magnetic field. It is almost certainly true that the magnetic field twists into and out of the line of sight in the various positions along the anomalous-velocity filaments analyzed here. A small twist could preferentially trigger the Hα signature (Figs. 2a & 3a) or the $C^+$, $N^+$ & $O^+$ signature (Figs. 2b & 3b).

## 4. CONCLUSIONS

The analysis of Verschuur & Schmelz (2022a) cleared the major obstacle preventing further investigations into the role the CIV effect may or may not be playing in interstellar space. It also established the CIV effect as a viable mechanism responsible for the widespread 34 km s$^{-1}$ feature in galactic λ21-cm HI profiles. Here, we build on results from the plasma physics literature that attributed the 25 km s$^{-1}$ feature to Hα (Verschuur & Schmelz 2022b).

Gaussian decomposition of HI4PI profiles was done for two regions along high-velocity Complex M, the first with corresponding Hα emission and the second without. In the former, we detected the same CIV signature found by Verschuur & Schmelz (2022b) − a line width of 25 km s$^{-1}$. In the latter, we found a statistically different CIV signature at 20 km s$^{-1}$ that we attribute to CNO$^+$.

We also discuss for the first time the different CIV signatures found in low-velocity and anomalous-velocity gas. In low-velocity gas, the CIV effect ionizes the atoms and produces the signatures of He (34 km s$^{-1}$), CNO (13.4 km s$^{-1}$), and the metals (around 6 km s$^{-1}$). In anomalous-velocity gas, the CIV effect ionizes the ions and produces the signatures of CNO$^+$ (20 km s$^{-1}$) and/or Hα (25.5 km s$^{-1}$), as well as the metals signatures around 6 km s$^{-1}$.

These distinct CIV signatures indicate that the physical conditions are different in each regime. The particles in the low-velocity gas cloud are initially neutral, the temperatures can be cold (50-100 K), and the abundances can be cosmic. The densities need to be high enough to impose regular recombination. The anomalous-velocity gas, on the other hand, initially contains a significant population of ionized and excited particles. The temperature and abundance values may be similar to the values that characterize low-density gas, but the densities must be low enough so that the plasma is essentially collision-less, similar to the conditions in the solar wind.

Finally, because the CIV is triggered as the particles reach a critical velocity normal to the magnetic field, it is important to learn more about the field orientation. Although observatories like Planck and SOFIA can give us the plane-of-sky component, neither observes cold interstellar space with the resolution required to constrain these results. Just as the Space Age revolutionized our understanding of interplanetary magnetic fields as magnetometers spread through the solar system, astrophysics requires an equivalent escalation to gain deeper insights in the physics of the interstellar medium.

## 5. ACKNOWLEDGMENTS
We are grateful to B. Winkel for sending us the EBHIS data, and to N. Brickhouse for useful advice.

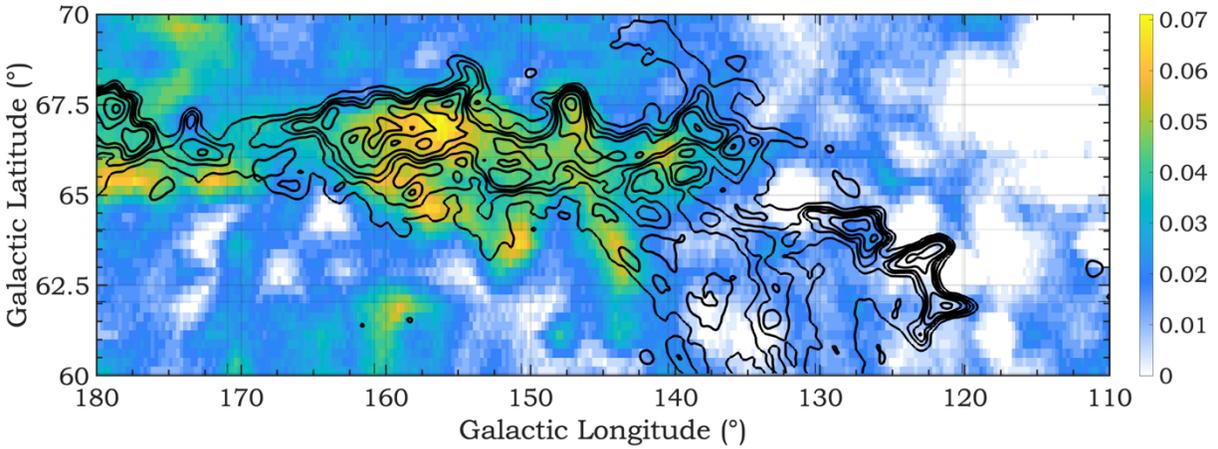

Figure 1. HI4PI map of l versus b for the region of Complex M that motivated this study. HI contours are superposed on the Hα image in color (legend in Rayleighs), both at the velocity of −80 km s$^{-1}$ with a 10 km s$^{-1}$ wide band. HI contours are 2:8 @2, 12, 20, 30 K.

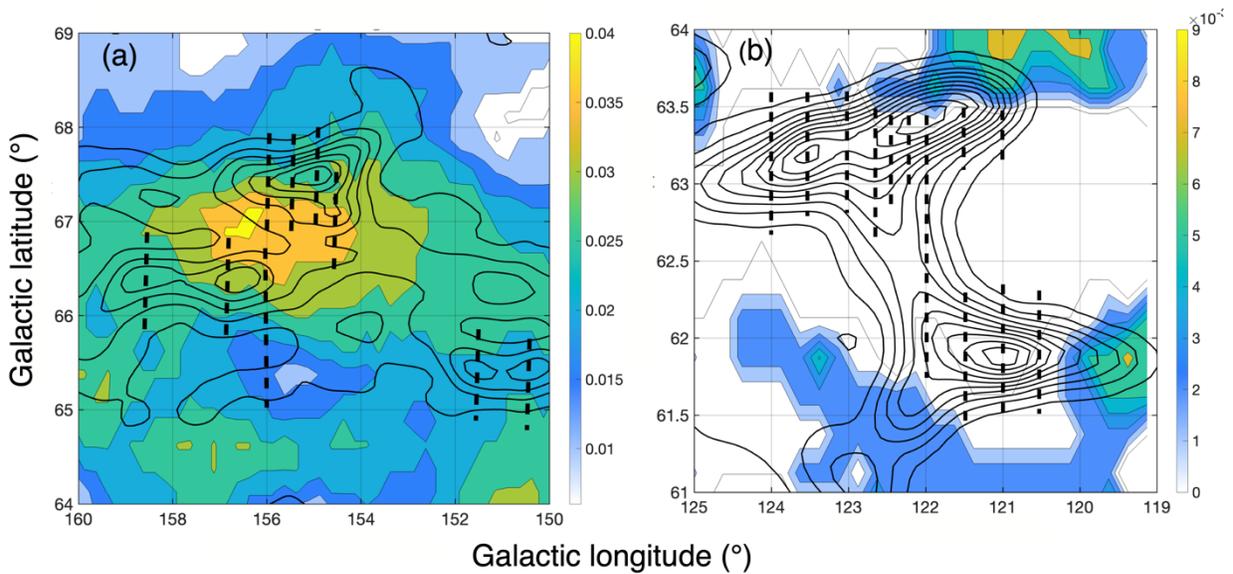

Figure 2. Close ups of the regions outlined in Fig. 1 with HI4PI contours superposed on the Hα image in color, both with a 4 km s$^{-1}$ wide band. Legends are in Rayleighs. (a) HI at -80 km s$^{-1}$ on the "Hα region" and (b) HI at -80 km s$^{-1}$ the "non-Hα region". The vertical dashed lines in each panel indicate the axes along which of HI profile at 0.°2 intervals were sampled. HI contour levels in (a) are from 0 to 12 K every 2 K and in (b) from 0 to 10 K in 1 K intervals.

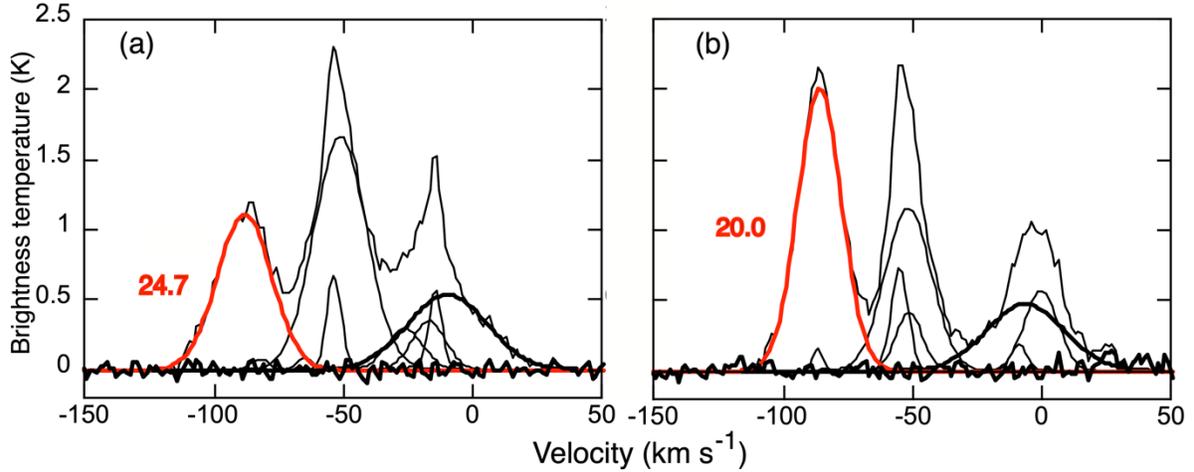

Figure 3. A typical spectrum for (a) the Hα region at l, b = (157.°1, 66.°0), and (b) the non-Hα region at l, b = (124.°0, 63.°1). Although the overall profiles appear complex, it is only the relatively isolated, left-most component (red) that is the target of the present study. The numbers beside these features are their line widths in km s$^{-1}$. The components at low velocities indicated by a thick line are members of the 34 km s$^{-1}$ family (see text).

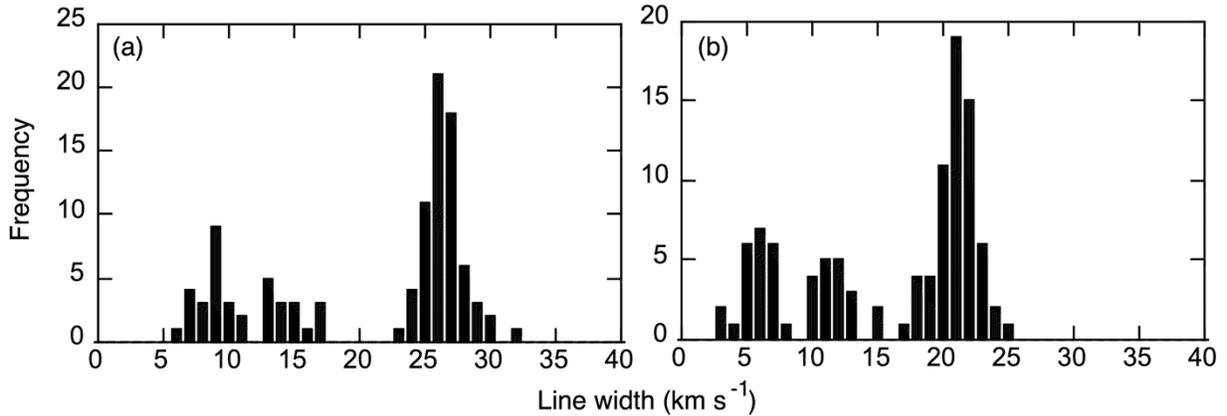

Figure 4. Line-width histograms for components with center velocities between -70 and -90 km s$^{-1}$ for the (a) Hα region and (b) non-Hα region, respectively. Each distribution shows the expected narrow components, but the new and most interesting feature is the dominant, wider peak in panel (a) at 25.9 ± 1.6 km s$^{-1}$, the CIV signature for Hα. In contrast, panel (b) shows a peak at 20.7 ± 1.5 km s$^{-1}$, the CIV signature of $CNO^+$.

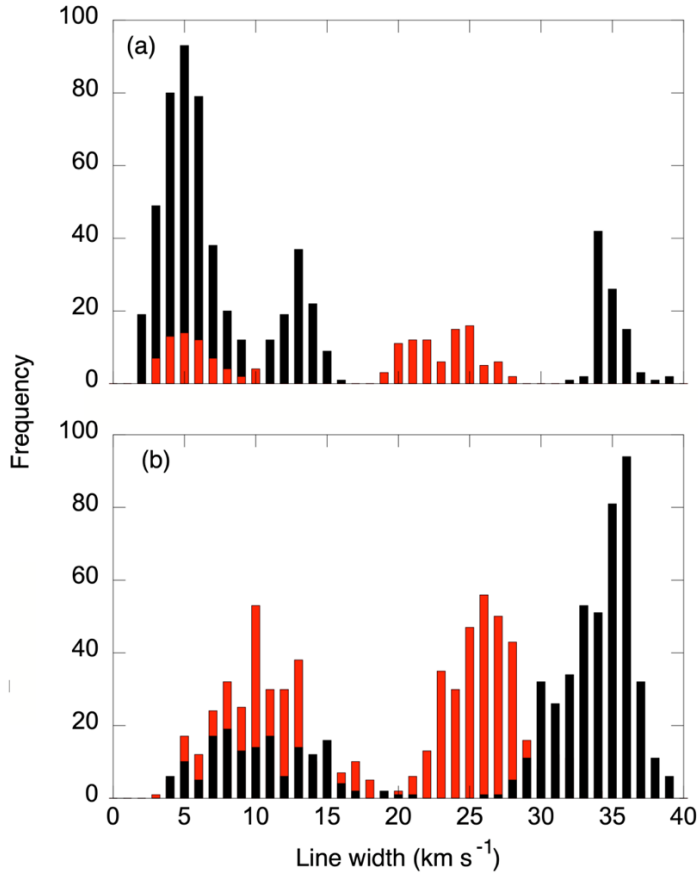

Figure 5. Gaussian line width distributions for (a) results from Verschuur et al. (2021) based on the mapping of a long, straight filament in the southern Galactic skies, and (b) previously unpublished results for a segment of the high-velocity cloud, MI (see text). Dark bars depict the line-width signatures of the low-velocity (-25 to +25 km s$^{-1}$) gas and red bars show those of the anomalous-velocity gas. The dark bars show peaks at approximately 34, 14, and 6 km s$^{-1}$, the CIV signatures of helium, CNO, and metals, respectively. (b) The red bars show clusters at 21 and 25 km s$^{-1}$, which represent the CIV signatures of Hα and CNO$^+$, respectively. Relatively weak associated Hα emission is present in both areas, see text.